\documentclass[12pt,twocolumn]{WileyMSP-template}
\usepackage{amsmath,amssymb}
\usepackage[square,super,sort&compress,comma]{natbib}
\usepackage{microtype}
\usepackage{xcolor}

\title{Magnon polaritons in a van der Waals ferromagnet coupled to a superconducting resonator}

\begin{document}
\fontsize{11}{12.5}\selectfont
\twocolumn[
\maketitle
\\
\author{
Alvaro Bermejillo-Seco,
Luuk J. van der Goot,
Matteo Arfini,
Yaroslav M. Blanter,
Gary A. Steele,
Herre S.J. van der Zant
\\[1ex]
Kavli Institute of Nanoscience, Delft University of Technology Lorentzweg 1, 2628 CJ Delft, Netherlands
}

\begin{abstract}
Achieving magnon–photon hybridization in the microwave regime is essential for integrating magnetic excitations with superconducting circuits. While this has been extensively demonstrated in bulk magnetic systems, realizing it in two-dimensional van der Waals materials remains challenging due to their reduced magnetic volume and increased dissipation. Here, magnon–photon hybridization is observed in exfoliated flakes of the van der Waals ferromagnet Cr$_2$Ge$_2$Te$_6$, with thicknesses down to 30 nm. The resulting magnon polaritons—hybrid excitations of cavity photons and magnons—are evidenced by reproducible avoided crossings across six devices, enabled by a low-impedance superconducting resonator design. The coupling strength follows the expected square-root dependence on thickness, and extrapolation of this scaling indicates that hybridization in the monolayer limit is within reach.
\end{abstract}
\vspace{0.5cm}
Corresponding authors: a.bermejilloseco@tudelft.nl, h.s.j.vanderzant@tudelft.nl
\vspace{0.7cm}
]

\section{Introduction}

Coupling different physical excitations provides a route to combine the advantages of distinct material platforms. Hybrid systems based on photons, phonons, and spins have therefore attracted significant attention in recent years. \cite{cavity_optomechanics2014, hybrid_systems2015,hybrid_systems2020,cavitymagnonics2022, cavity_magnonics2_2022, Engelhardt2022} In magnetic materials, the collective spin excitations known as magnons can interact coherently with microwave photons through the Zeeman interaction. \cite{Huebl2013, Tabuchi2014, zhang2014, Goryachev2014, Bai2015} Such magnon–photon interactions have enabled the exploration of cavity magnonics and have been proposed as building blocks for microwave signal processing and quantum transduction between different frequency domains. \cite{cavitymagnonics2022,cavity_magnonics2_2022,Tabuchi2015,Zhang2016,Lachance2017,Wang2018,Lachance2020}

Strong magnon–photon coupling was first realized using bulk yttrium iron garnet (YIG) crystals coupled to microwave cavities. \cite{Huebl2013, Tabuchi2014, zhang2014, Goryachev2014, Bai2015} Since then, considerable effort has been devoted to reducing the magnet size and improving integration with planar superconducting circuits.\cite{Mckenzie2019,Hou2019,Li2019,Golovchanskiy2021,Ghirri2023,Ghirri2026} Van der Waals (vdW) magnets offer an appealing materials platform in this context. Their layered structure allows exfoliation down to nanometer thickness and facilitates integration with other two-dimensional materials in heterostructures combining magnetic, electronic, optical, and mechanical functionalities.\cite{Huang2017,Burch2018,Yang2021,Gibertini2019,Grubisic_2025} However, the small magnetic volume of exfoliated flakes strongly suppresses the collective magnon–photon coupling, and vdW magnets typically exhibit significantly higher frequencies and larger magnetic losses than YIG.\cite{Wang2026} As a result, achieving magnon–photon hybridization in thin flakes remains challenging.\cite{Zollitsch2023} Overcoming this limitation would enable integrated magnonic devices based on vdW magnets and provide a microwave probe of magnon spectra and damping as these materials approach the two-dimensional limit.

We demonstrate magnon–photon hybridization in exfoliated flakes of the van der Waals ferromagnet Cr$_2$Ge$_2$Te$_6$ (CGT) with thicknesses down to 30 nm. Hybridization is observed across multiple devices by coupling the flakes to a planar superconducting microwave resonator designed to concentrate the microwave magnetic field within a small mode volume. The resulting interaction produces clear avoided crossings between the cavity and magnon modes, evidencing hybridization in thin vdW magnets. 

\section{Results and Discussion}
\begin{figure*}[h!]
    \centering
    \includegraphics[width=0.75\linewidth]{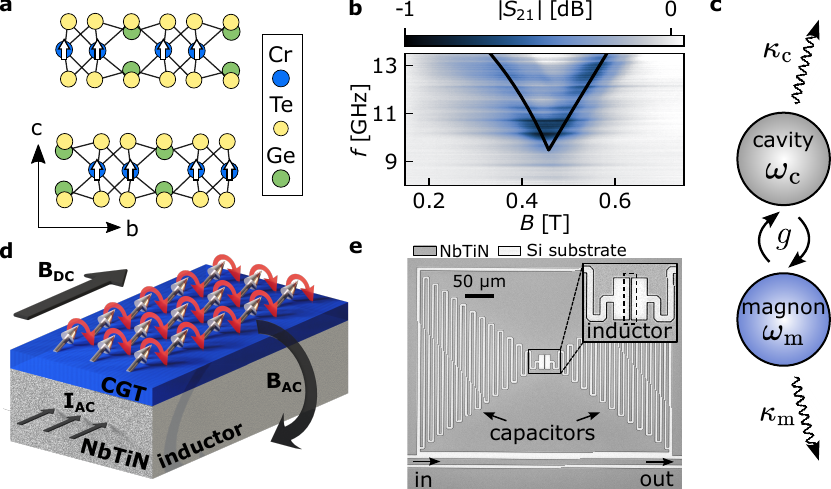}
    \caption{\textbf{a} Crystal structure of CGT, showing Cr spins aligned along the crystallographic c-axis due to magnetic anisotropy.
    \textbf{b} Spinwave absorption spectrum of a bulk CGT crystal measured on a 50 $\Omega$ feedline as a function of in-plane magnetic field at 2 K. The black line is a theoretical calculation (see Section \textcolor{blue}{S2}, Supporting Information).
    \textbf{c} Diagram of a magnon mode with frequency $\omega_\text{m}$ coupled to a cavity photon mode with frequency $\omega_\text{c}$ at rate $g$. The respective dissipation rates are $\kappa_\text{m}$ and $\kappa_\text{c}$.
    \textbf{d} Schematic of the coupling mechanism. An AC current at the cavity frequency flows through the NbTiN inductor, generating an oscillating magnetic field that couples to the uniform magnon mode in the CGT flake placed on top. A DC in-plane magnetic field tunes the magnon frequency.   
    \textbf{e} Microscope picture of the low-impedance superconducting resonator. The arrows indicate the input and output of the 50 $\Omega$ feedline to which it is side-coupled. The interdigitated capacitors are located on either side of the inductor; the inset shows a magnified view of this region, where the inductor is indicated by a dashed rectangle. The gray regions correspond to the NbTiN film, while the white areas indicate where the NbTiN has been etched, exposing the underlying Si substrate.
    }
    \label{fig:1}
\end{figure*}

Below its Curie temperature of 62~K~\cite{Gong2017,Khan2019}, CGT is a ferromagnet with out-of-plane easy-axis anisotropy whose crystal structure is shown in \textbf{Figure~\ref{fig:1}a}. When a DC magnetic field is applied in the plane, the spins cant toward the field, and the magnon frequency decreases, reaching a minimum at a critical field of approximately 450 mT. This critical field is determined by the crossing of two magnon branches and is discussed in the Supporting Information, Section \textcolor{blue}{S2}.\cite{SI_supp}. Here, we focus on the fields above this value, where the magnetization is fully aligned in-plane and magnetic domains are absent. In this regime, the magnon frequency increases following the Kittel relation\cite{Shen2021} 
\begin{equation}
    \omega_\text{m}/2\pi = \gamma\mu_0\sqrt{H_\text{DC}\left(H_\text{DC}+M_\text{s} - \frac{2K_\text{u}}{\mu_0M_\text{s}}\right)}.
    \label{eq:omega0}
\end{equation}
Here, $H_\text{DC}$ is the external magnetic field, $\gamma$ the gyromagnetic ratio, $M_\text{s}$ the saturation magnetisation, and $K_\text{u}$ the magnetic anisotropy. We measure the ferromagnetic resonance of a bulk CGT crystal at 2 K (\textbf{Figure~\ref{fig:1}b}) and find good agreement with the calculated spin-wave spectrum and literature values for $\mu_0M_\text{s} = 211$ mT, $K_\text{u} = 3.84 \times 10^4$ Jm$^{-3}$, and $\gamma = 30.55$ GHzT$^{-1}$\cite{Zollitsch2019}. These measurements serve as a reference for selecting the magnetic-field range used in the experiments presented below, corresponding to the fully in-plane magnetized regime. 

The hybrid system consists of a cavity photon mode with frequency $\omega_\text{c}$ and loss rate $\kappa_\text{c}$, coupled to a magnon mode with frequency $\omega_\text{m}$ and loss rate $\kappa_\text{m}$ at rate $g$, as schematically shown in \textbf{Figure~\ref{fig:1}c}. The coupling mechanism is illustrated in \textbf{Figure~\ref{fig:1}d}: an AC current in the superconductor generates an oscillating magnetic field that exerts a torque on the spins in the magnet, thereby driving the magnon mode. The coupling strength between the cavity and magnon mode is given by
\begin{equation}
    g = \frac{\eta}{2}\gamma\sqrt{\hbar \mu_0 \omega_c}\sqrt{\rho_s s},
    \label{eq:coupling}
\end{equation}
where $\eta$ is the overlap factor between the resonator microwave magnetic field and the magnon mode, $\rho_s$ is the spin density, and $s$ is the spin number per site.\cite{Soykal2010,Mckenzie2019,cavitymagnonics2022} The interplay between the coupling strength $g$ and the dissipation rates of the magnon ($\kappa_\text{m}$) and cavity ($\kappa_\text{c}$) defines different regimes of operation. Hybridization, observed as an avoided crossing in the spectrum, occurs when $ 4g > |\kappa_\text{m} - \kappa_\text{c}|$.\cite{cavity_optomechanics2014,Rodriguez2016} The system is typically considered to be in the strong-coupling regime when $g \gtrsim \kappa_\text{m}, \kappa_\text{c}$, such that coherent exchange between the two systems dominates over dissipation. As an additional figure of merit, the cooperativity $\mathcal{C}=4g^2/(\kappa_\text{m}\kappa_\text{c})$ is often used to quantify the ratio of coherent coupling to losses.

For a magnet in a spatially homogeneous microwave field, $\eta$ scales as $\sqrt{N}$ with the number of spins, which for a vdW flake of fixed lateral area reduces to a $\sqrt{t}$ dependence on thickness. This scaling presents a fundamental challenge for ultrathin flakes, where the small spin ensemble suppresses $g$, making the design of resonators that maximize $\eta$ essential. To address this, we employ a low-impedance, lumped-element resonator~\cite{Eichler2017,Mckenzie2019} patterned in a 100~nm thick NbTiN film (\textbf{Figure~\ref{fig:1}e}), consisting of two interdigitated capacitors connected by a narrow inductor. In this geometry, the microwave current --- and therefore the microwave magnetic field --- is strongly concentrated within the inductor volume. Placing a CGT flake directly on top of this inductor maximizes the spatial overlap between the resonator field and the spin ensemble, thereby boosting $\eta$. The resonator is side-coupled to a 50~$\Omega$ transmission line for microwave readout. Further details on the resonator design and its magnetic-field resilience are provided in the Supporting Information, Section \textcolor{blue}{S1}.\cite{SI_supp}

\begin{figure*}[h]
    \centering
    \includegraphics[width=0.85\linewidth]{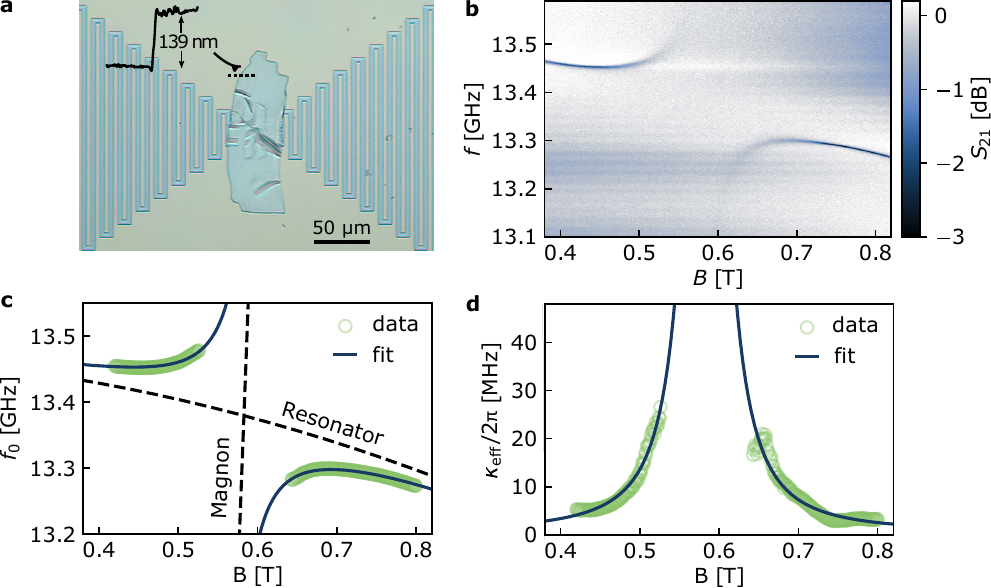}
    \caption{\textbf{a} Microscope picture of a CGT flake transferred on top of the inductor wire of a low-impedance resonator (device A). The flake is 139 nm thick, as deduced from atomic force microscopy (inset).
    \textbf{b} Background-subtracted absorption spectrum of the device shown in \textbf{a} at 2K as a function of in-plane magnetic field showing an avoided crossing.
    \textbf{c},\textbf{d} Analysis of the resonance frequency (\textbf{c}) and linewidth (\textbf{d}) as a function of the in-plane magnetic field. The data points are obtained by fitting Lorentzian line shapes to the spectra in \textbf{b} and Equation \ref{omegapm} is fitted to them. The fitted resonance frequencies of the uncoupled modes are shown as dashed lines. The extracted fit parameters are $g/2\pi=386$ MHz, $\kappa_\text{c}/2\pi=1.39$ MHz, and $\kappa_\text{m}/2\pi=650$ MHz.
    }
    \label{fig:2}
\end{figure*}

An experimental realization of a CGT flake integrated with a superconducting resonator is shown in \textbf{Figure} \ref{fig:2}\textbf{a} (device A). The 139 nm thick CGT flake is stamped onto a resonator with a resonance frequency of 13.9 GHz and an inductor that is 1 \textmu m wide and 50 \textmu m long. The flake was chosen to be large enough to fully cover the inductor in order to maximize $\eta$.  The system is probed by measuring the microwave transmission, $S_{21}$, through the transmission line to which it is coupled, as a function of frequency and external in-plane magnetic field. The measurements are done at a temperature of 2 K, well below the superconducting transition of NbTNi.\cite{Thoen2017} The resulting spectrum is shown in \textbf{Figure} \ref{fig:2}\textbf{b}. A clear avoided crossing between the superconducting resonator mode and the magnon mode is observed, indicating magnon--photon hybridization.

To quantitatively describe the hybridized modes, we model the system as a coupled magnon–photon oscillator.\cite{cavity_optomechanics2014,Rodriguez2016} The complex eigenfrequencies ($\omega_\pm$) are then given by
\begin{equation}\label{omegapm}
\omega_\pm = \tilde\omega \pm \sqrt{\Delta^2/4 + g^2},
\end{equation}
where $\tilde\omega=(\omega_\text{m}+\omega_\text{c})/2-i(\kappa_\text{m}+\kappa_\text{c})/4$ is the mean complex frequency of the uncoupled modes and $\Delta = (\omega_\text{c}-\omega_\text{m})-i(\kappa_\text{c}-\kappa_\text{m})/2$ their complex detuning. The coupling modifies the spectrum through the square-root term, which provides the basis for the hybridization condition introduced above. On resonance, when $4g \leq |\kappa_\text{m}-\kappa_\text{c}|$, the square root remains imaginary, and the interaction mainly results in additional damping without producing two distinct modes. In contrast, when $4g > |\kappa_\text{m}-\kappa_\text{c}|$, the square root acquires a real component and the spectrum splits into two branches with different real frequencies, giving rise to the avoided crossing observed in the experiment. In our devices, the cavity loss rate is negligibly small compared to the magnon loss rate ($\kappa_\text{c} \ll \kappa_\text{m}$), such that the condition for hybridization reduces to $g > \kappa_\text{m}/4$.

The real and imaginary parts of $\omega_\pm$ correspond to the resonance frequency and loss rate of each hybridized mode, respectively. These quantities are extracted from the spectrum in \textbf{Figure} \ref{fig:2}\textbf{b} by fitting the resonances in the $S_{21}$ response at each magnetic field to obtain the mode frequency and linewidth (see Supporting Information, Section \textcolor{blue}{S1}, for details). The extracted resonance frequencies and loss rates are plotted in \textbf{Figure} \ref{fig:2}\textbf{c,d}. We then perform a global fit of these data using Equation \ref{omegapm}, assuming the Kittel relation for the magnon frequency $\omega_\text{m}$ (Equation \ref{eq:omega0}) and a quadratic magnetic-field dependence for the cavity frequency $\omega_c$. For more details about the fitting, see Supporting Information, Section \textcolor{blue}{S4}. The model reproduces the data well and yields $g/2\pi=386$ MHz, $\kappa_\text{c}/2\pi=1.39$ MHz, and $\kappa_\text{m}/2\pi=650$ MHz. These values confirm that $\kappa_\text{m} \gg \kappa_c$ and that the hybridization condition $g>\kappa_\text{m}/4$ is satisfied.

\begin{figure}[h!]
    \centering
    \includegraphics[width=0.9\linewidth]{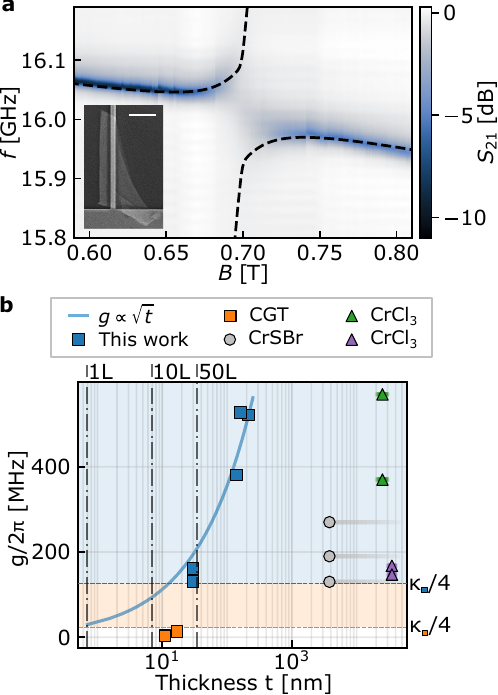}
    \caption{\textbf{a} Background-subtracted absorption spectrum of 30nm flake (device B) measured at 2 K as a function of in-plane magnetic field showing an avoided crossing. The dashed line shows the fitted resonance frequencies. The inset displays an SEM image of the flake on the inductor (scale bar: 5 \textmu m). 
    \textbf{b} Overview of reported couplings between microwave cavities and van der Waals magnets as a function of flake thickness. Five devices from this work are shown in blue squares and are fitted to a square root dependence. The threshold for observing hybridisation is indicated by the blue dashed line corresponding to the losses measured in this work and the orange dashed line corresponding to the lowest losses reported in Ref. \cite{Zollitsch2023}. Vertical dash-dot lines mark the thicknesses corresponding to 1, 10, and 50 layers, assuming a monolayer thickness of 0.7 nm. Data in the weak-coupling regime for CGT are taken from \cite{Zollitsch2023} (orange squares), while data showing hybridization in the bulk regime are taken for CrSBr from  \cite{Tang2025} (grey circles) and for CrCl$_3$ from \cite{1zhang2021,2Zhang2021} (purple triangles) and  \cite{Mandal2020} (green triangles). The thickness for the CrSBr flakes is not explicitly reported, but quoted to be higher than 3.8 \textmu m.
    } 
    \label{fig:3}
\end{figure}

Magnon-photon hybridization is reproduced in six devices (A-F) with CGT flakes of different thicknesses. A summary of the devices is shown in \textbf{Table} \ref{tab:1}; the raw data and fits of all devices can be found in the Supporting Information, Section \textcolor{blue}{S4}, as well as AFM scans.\cite{SI_supp}. For all devices, the condition $g>\kappa_\text{m}/4$ is satisfied, consistent with the avoided crossings observed in the spectra. The coupling decreases with thickness as expected, with 30 nm being the thinnest devices (B and C). The spectrum and fit of one of the devices with a 30 nm flake on a 25 \textmu m long inductor (device B) is shown in \textbf{Figure} \ref{fig:3}\textbf{a}. The cooperativity in our devices ranges from hundreds in the thickest flakes down to 5.9 in the thinnest flake. All the cooperativities exceed unity, indicating that the coupling exceeds the effective dissipation scale; however, the devices are not in the strong coupling regime according to the condition, $g\geq \kappa_\text{c},\kappa_\text{m}$, due to the large magnon dissipation rates. 

\begin{table*}[h!]
\centering
\renewcommand{\arraystretch}{1.25}
\setlength{\tabcolsep}{8pt}
\caption{Summary of device characteristics and fitted magnon--photon coupling parameters. Boxes filled with -- indicate that the information is unknown or the fit was considered unreliable. The raw spectra, fits, and AFM scans for all devices are provided in the Supporting Information, Section \textcolor{blue}{S4}.\cite{SI_supp} The stamping methods are detailed in the Experimental Section below.}
\begin{tabular}{cccccccccc}
\hline
Device & Inductor & Flake  & Flake & $g/2\pi$ & $\kappa_\text{m}/2\pi$ & $\kappa_\text{c}/2\pi$ & $\mathcal{C}$ & Stamping \\
 & Length [\textmu m] & Thickness [nm] & Volume [\textmu m$^3$] & [MHz] & [MHz] & [MHz] & $4g^2/\kappa_\text{c}\kappa_\text{m}$ & Method \\
\hline
A & 50 & 139 & 906.3 & 381 & 586 & 2.3 & 426.4 & PC \\
B & 25 & 30 & 3.4 & 161 & 510 & 29.0 & 5.9 & PDMS \\
C & 25 & 30 & 2.0 & 131 & 453 & 22.9 & 6.7 & PDMS \\
D & 25 & 162 & 29.8 & 527 & 725 & 17.5 & 87.2 & PDMS \\
E & 50 & 214 & 219.4 & 522 & -- & -- & -- & PC \\
F & 25 & -- & -- & 710 & -- & -- & -- & PDMS \\
\hline
\end{tabular}
\label{tab:1}
\end{table*}

An important question is what is the minimum thickness at which magnon–photon hybridization can be observed. We address this in \textbf{Figure} \ref{fig:3}\textbf{b}, where the coupling strengths extracted from our devices are plotted as a function of flake thickness and fitted to the square-root dependence predicted by Eq.~\ref{eq:coupling}. The fit agrees well with the experimental data and, combined with the measured magnon loss rates, allows us to estimate the thickness limit for hybridization. For a typical magnon linewidth in our devices of approximately 500 MHz, the hybridization threshold is reached slightly above 10 nm, where the fit intersects the blue horizontal dashed line at $\kappa_\text{m}/4$. If the magnon losses were reduced to values comparable to the best reported in Ref.~\cite{Zollitsch2023} (around 100 MHz), and the square root dependence holds when extrapolated to thinner flakes, hybridization would remain possible with this resonator design even in the monolayer limit, as indicated by the orange dashed line. \textbf{Figure} \ref{fig:3}\textbf{b} also includes data from previous studies, highlighting the improvement achieved in this work: hybridization is observed in flakes that are two to three orders of magnitude thinner than those used in experiments with vdW layered antiferromagnets (circles and triangles in \textbf{Figure} \ref{fig:3}\textbf{b}) \cite{1zhang2021,2Zhang2021,Mandal2020}, while the coupling strength in CGT is increased by roughly two orders of magnitude compared to previous reports (orange squares in \textbf{Figure} \ref{fig:3}\textbf{b}) \cite{Zollitsch2023}.

These considerations clarify the main challenges for reaching the monolayer limit, with a thickness of 0.7 nm.\cite{Gong2017} A key requirement is maximizing the lateral overlap between the magnetic flake and the resonator inductor, since the coupling strength depends directly on the spatial overlap between the microwave magnetic field and the spin ensemble. This implies that exfoliation techniques capable of producing large-area monolayers are essential. In addition, CGT is known to degrade in air by forming a thin oxide layer. Consequently, device fabrication must either be performed entirely under inert conditions or the flakes must be protected, for example, by capping or encapsulation with hBN. The relatively large damping rates observed in the present work may partly originate from this effect, as the flakes were exfoliated and transferred outside a glovebox and were exposed to air for approximately one to two hours before measurement.

Additionally, a possible origin of the observed large magnon linewidths is the presence of multiple modes, as reported in Ref.~\cite{Zollitsch2023}. We observe faint diagonal lines at the center of the avoided crossing in devices C and D, consistent with additional modes. While the avoided crossings are well described by a single effective magnon mode, these additional modes likely contribute to an increased apparent linewidth. Further discussion of these features is provided in the Supporting Information, Section~\textcolor{blue}{5}.

\section{Conclusions}

In summary, we demonstrate magnon--photon hybridization between superconducting microwave resonators and exfoliated flakes of the van der Waals magnet Cr$_2$Ge$_2$Te$_6$. By employing a low-impedance lumped-element resonator that concentrates the microwave magnetic field in a small mode volume, we achieve coupling strengths in the hundreds of MHz range and observe clear avoided crossings for flakes as thin as 30 nm. The measured coupling follows the expected square-root dependence on the flake thickness, highlighting the importance of resonator designs that maximize magnetic field overlap when working with small magnetic volumes. Extrapolation of this scaling suggests that hybridization could be achieved in substantially thinner flakes, potentially approaching the monolayer limit, if magnon losses are further reduced and large-area monolayers become available.

\section{Experimental Section}

\threesubsection{Resonator fabrication} The device is fabricated on $10\times10$ mm 525 \textmu m thick high-resistivity silicon chips with 100 nm of niobium titanium nitride (NbTiN). The NbTiN is deposited by the Dutch Institute for Space Research (SRON) following the procedure described in \cite{Thoen2017}. The device is fabricated by spinning a (positive) e-beam resist layer (AR-P 6200.18, 4000 rpm) and patterning the CPW, resonator, and feedlines using electron-beam lithography. After development (Pentylacetate 60s, O-xylene 5s, IPA 60s), the NbTiN is removed using a CF$_4$ reactive ion etching step, followed by an in-situ oxygen plasma descum. Finally, the remaining resist is stripped in Dimethylformamide (DMF).

\threesubsection{Stamping procedures} Two different methods have been used to stamp flakes onto the superconducting resonator inductors. The exfoliation and transfer of multi-layer CGT flakes were done using a combination of polydimethylsiloxane (PDMS) \cite{Castellanos2014} and polycarbonate (PC) \cite{Zomer2014} transfer methods. First, CGT crystals were exfoliated onto the PDMS through scotch tape. Selected flakes were then transferred to the inductor of the superconducting resonator. For devices A and E, the CGT crystals were exfoliated using scotch tape directly on Si/SiO2 substrates. Selected flakes are then transferred to the inductor using PC on PDMS. For device F, the PDMS was left behind covering the flake to act as a protection layer, which prevented thickness identification.

The PC method leads to very high yield in flake transferring, but requires heating to 180 °C, which accelerates the oxidation of CGT. The PDMS transfer can be done at room temperature and requires good control of the angle and speed while releasing the flake. This added difficulty is due to the lack of full contact of the flake with the substrate, since it hangs into the trench on either side of the inductor.

\threesubsection{Wirebonding} After the transfer of the flake onto the superconducting device, the chip was mounted and wire-bonded onto a printed circuit board equipped with two RF lines. A trade-off exists between the time required for wire bonding and the resulting electrical performance. To ensure homogeneous grounding of the chip at microwave frequencies, ground bonds were placed with a spacing of $\leq$ 0.5 mm. This dense grounding suppresses parasitic resonances and impedance inhomogeneities, resulting in a cleaner RF spectrum up to 18 GHz. The RF signal pads were connected using two to three wire bonds per pad to reduce inductance and improve impedance matching. Using our wire-bonding setup, this bonding scheme can be reliably implemented within 30 minutes per device.

\threesubsection{Measurement setup} The devices were measured in an Attodry 2100 cryostat with a base temperature of 1.67 K and equipped with a magnet providing an in-plane magnetic field of up to 9 T. The measurement insert, supplied by VASTA, provided two RF lines with a bandwidth up to 18 GHz. To ensure operation of the superconducting resonator in the linear regime, fixed attenuators of 20 dB at room temperature and 20 dB at the cryogenic stage were installed on the input line before the PCB. The transmitted signal was amplified before detection by the vector network analyzer using a Mini-Circuits ZX60-183-S+ amplifier, providing 24 dB of gain over the 6–18 GHz frequency range.

\medskip
\textbf{Acknowledgements} \par 
 We acknowledge T.S. Ghiasi, S. Deve, and M. Villiers for helpful discussions and Y. Yang for help with the experimental setup. A.B.-S., H.S.J.vdZ. and Y.M.B. acknowledge support by the Dutch Research Council (NWO) under the project ``Ronde Open Competitie XL" (file number OCENW.XL21.XL21.058). M.A. and G.A.S. acknowledge support by the Dutch Research Council (NWO) under the project number VI.C.212.087 of the research programme VICI round 2021.

\medskip

\bibliographystyle{MSP}
\bibliography{main}

\end{document}


\maketitle
\\
\author{
Alvaro Bermejillo-Seco,
Luuk J. van der Goot,
Matteo Arfini,
Yaroslav M. Blanter,
Gary A. Steele,
Herre S.J. van der Zant
\\[1ex]
Kavli Institute of Nanoscience, Delft University of Technology Lorentzweg 1, 2628 CJ Delft, Netherlands
\\[1ex]
Corresponding authors: a.bermejilloseco@tudelft.nl, h.s.j.vanderzant@tudelft.nl
}
\vspace{0.3cm}

\tableofcontents


\section{Superconducting resonator design and characterisation}

The general design of the resonators used in this work follows Refs.~\cite{Mckenzie2019}. The geometry is based on a lumped-element resonator consisting of a narrow inductive wire, where the microwave magnetic field is strongly confined. In this work, inductors with dimensions $50\,\mu$m $\times$ $1\,\mu$m and $25\,\mu$m $\times$ $1\,\mu$m were used. The small width is chosen to enhance the inductance relative to the rest of the circuit, while the length is selected to match the lateral size of available flakes, maximizing the spatial overlap between the magnetic field and the magnetic material.

On either side of the inductor, interdigitated capacitors with a finger width of $4\,\mu$m and varying finger length and number are used to set the capacitance. This allows tuning of the resonance frequency, which in our case was designed to be in the range of 13–16 GHz. These frequencies are chosen such that the resonator mode intersects the magnon dispersion above $\sim 450$ mT, where the magnetization is aligned in-plane, as discussed in the main text.

The resonators are designed using the Ansys HFSS eigenmode solver. An important parameter is the kinetic inductance of the NbTiN thin films. Due to the disordered nature of the superconductor and its thickness of 100 nm, the kinetic inductance provides a significant contribution, particularly in the narrow inductor, and strongly affects the resonance frequency. To estimate this contribution, a reference resonator was fabricated and measured without including kinetic inductance in the simulation. By comparing simulated and measured resonance frequencies, a kinetic inductance of $1.7$ pH/square was extracted. Including this value in the simulations allows prediction of the resonance frequency within a few percent accuracy. The simulations also confirm that the microwave magnetic field is confined to a small region around the inductor.

The resilience of the resonators to in-plane magnetic fields is characterized by measuring the transmission $S_{21}$ as a function of frequency and magnetic field. Each trace is fitted using a standard model for a side-coupled resonator:
\begin{equation}\label{S21}
   S_{21}(\omega) = 1 - \frac{\kappa_{\text{ext}} e^{i\theta}}{\,\kappa_{\text{ext}} + \kappa_{\text{int}} + 2 i\, (\omega-\omega_0)\,},
\end{equation}
where $\omega_0$ is the resonance frequency, $\kappa_{\text{int}}$ and $\kappa_{\text{ext}}$ are the internal and external loss rates, respectively, and $\theta$ accounts for a global phase offset. The total loss rate is given by $\kappa = \kappa_{\text{int}} + \kappa_{\text{ext}}$.

\begin{figure}
    \centering
    \includegraphics[width=\linewidth]{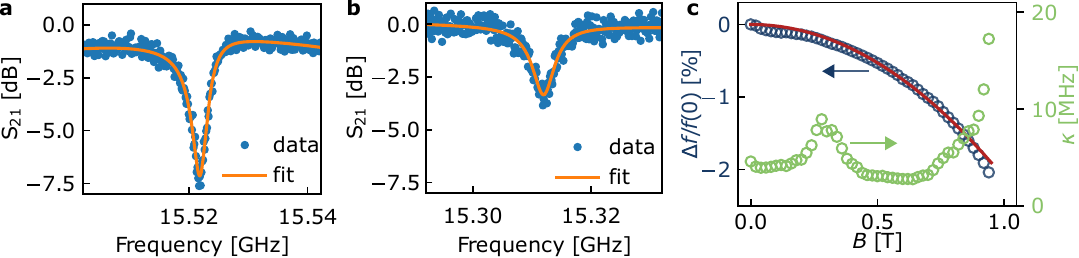}
    \caption{\textbf{a} Fit to a superconducting resonator at 2K at 0 T with a fit rendering $f_0=15.52$ GHz, $\kappa_{int}=2.14$ MHz, $\kappa_{ext}=2.44$ MHz, $\kappa_c=4.59$ MHz. \textbf{b} Fit to a superconducting resonator at 2K at 800 mT with a fit rendering $f_0=15.31$ GHz, $\kappa_{int}=3.79$ MHz, $\kappa_{ext}=1.74$ MHz, $\kappa_c=5.53$ MHz. \textbf{c} The blue dots (left axis) show the relative resonance frequency, normalized to its zero-field value of 15.52 GHz, as a function of magnetic field for a resonator without a flake. A fit to Equation~\ref{eq:quadratic} captures the expected quadratic dependence. The green dots (right axis) show the corresponding total linewidth as a function of magnetic field.  The blue and green arrows indicate which axis should be used to read each dataset, pointing to the left and right axes for the frequency and linewidth data, respectively.}
    \label{fig:SI1}
\end{figure}

\textbf{Figure}~\ref{fig:SI1}\textbf{a,b} show representative spectra at 0 mT and 800 mT, together with the corresponding fits, yielding the resonance frequency and loss rates. The evolution of these quantities as a function of in-plane magnetic field is shown in Figure ~\ref{fig:SI1}\textbf{c}. The resonance frequency follows a quadratic dependence on the magnetic field,
\begin{equation}\label{eq:quadratic}
    f_0(B) = f_0(0) - aB^2,
\end{equation}
which is attributed to an increase in kinetic inductance with magnetic field~\cite{Roy2025}. Accounting for this dependence is important for accurately modelling the avoided crossings discussed in the main text.

The total linewidth exhibits a peak around 300 mT, which is attributed to coupling to parasitic two-level systems, likely associated with defects in the silicon substrate~\cite{Zollitsch2019}. Outside this region, the linewidth remains below 10 MHz up to 900 mT, demonstrating that the resonators maintain low losses over the full field range relevant for the experiments.

\section{Magnon dispersion in bulk CGT}

\begin{figure}[h]
    \centering
    \includegraphics[width=0.65\linewidth]{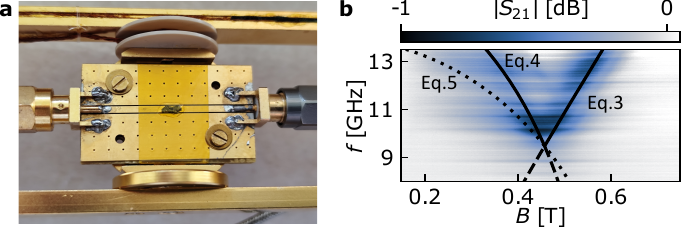}
    \caption{\textbf{a} Bulk CGT sample on a 50 $\Omega$ line fixed with kapton tape. \textbf{b} Background substracted transmission spectrum, $S_{21}$, of the sample shown in \textbf{a} with a calculation of the dispersion according to Equation \ref{eq:omegafmr}, \ref{eq:omegaphi0} and \ref{eq:omegaphi90} as indicated by the labels. }
    \label{fig:bulkCGT}
\end{figure}

The magnetic response of the sample is characterized by its ferromagnetic resonance (FMR) frequency. Measurements are performed by placing a bulk Cr$_2$Ge$_2$Te$_6$ (CGT) crystal on a 50~$\Omega$ transmission line (see Fig.~\ref{fig:bulkCGT}a) and measuring the microwave transmission, $S_{21}$, as a function of frequency and applied in-plane magnetic field. In the case of a uniformly magnetized state, achieved at sufficiently large in-plane magnetic fields, the resonance frequency is given by
\begin{equation}\label{eq:omegafmr}
    \omega_\text{m} = \gamma\mu_0\sqrt{H_\text{dc}\left(H_\text{dc}+M_s - H_\text{k}\right)},
\end{equation}
where $M_s$ is the saturation magnetization and $H_k = 2K_{u1}/(\mu_0 M_s)$ is the effective anisotropy field. The corresponding resonance is observed as a dip in the background-subtracted transmission spectrum shown in Fig.~\ref{fig:bulkCGT}b.

At lower magnetic fields, the magnetization is not fully aligned, and the sample forms a multidomain state~\cite{Vervelaki2024}. In this regime, the internal magnetic field is modified by the domain structure, leading to additional resonance modes. Following Ref.~\cite{shen_2021}, two characteristic modes can be identified depending on the orientation of the external field relative to the domain walls. For a field perpendicular to the domain walls, the resonance frequency is
\begin{equation}\label{eq:omegaphi0}
    \omega_\text{m}= \gamma\mu_0 \sqrt{H_\text{k}^2 + M_s H_\text{k} - H_\text{dc}^2},
\end{equation}
while for a field parallel to the domain walls, it becomes
\begin{equation}\label{eq:omegaphi90}
    \omega_\text{m}= \gamma \mu_0 \sqrt{H_\text{k}^2 + \frac{M_s H_\text{dc}^2}{H_\text{k}} - H_\text{dc}^2}.
\end{equation}

These expressions provide a quantitative description of the measured resonance frequencies, allowing the calculated dispersions to be overlaid on the experimental data in Fig.~\ref{fig:bulkCGT}b for comparison. The experimental data is best described by the mode corresponding to the field perpendicular to the domain walls. The parameters used in the equations are $\mu_0M_s = 211$ mT, $K_u = 3.84 \times 10^4$ Jm$^{-3}$, and $\gamma = 30.55$ GHzT$^{-1}$.\cite{Zollitsch2019} The resulting magnon spectrum follows Equation \ref{eq:omegaphi0} up to $\sim$450 mT, where it crosses Equation \ref{eq:omegafmr}. No evidence is found for magnons following Equation \ref{eq:omegaphi90}.

\section{Coupled magnon photon model}

A magnon mode with frequency $\omega_m$ and loss rate $\kappa_m$ (defined as its full width at half maximum), coupled to a cavity photon mode with frequency $\omega_c$ and loss rate $\kappa_c$ with strength $g$, can be described near resonance by the non-Hermitian Hamiltonian~\cite{cavity_optomechanics2014,Rodriguez2016}
\begin{align}
    |\mathbf{H}-\mathbf{I}\omega| = 0, \quad 
    \mathbf{H} = \begin{pmatrix}
    \omega_c - \frac{i\kappa_c}{2} & g \\
    g & \omega_m - \frac{i\kappa_m}{2}
    \end{pmatrix}.
    \label{Eq:matrix}
\end{align}

Solving Equation~\ref{Eq:matrix} yields the complex eigenfrequencies
\begin{align}\label{eq:omegapm}
\omega_\pm = \tilde\omega \pm \sqrt{\frac{\Delta^2}{4}+g^2},
\end{align}
where $\tilde\omega=(\omega_m+\omega_c)/2 - i(\kappa_m+\kappa_c)/4$ is the average complex frequency and $\Delta = (\omega_c-\omega_m) - i(\kappa_c-\kappa_m)/2$ is the complex detuning. The real part of $\omega_\pm$ gives the resonance frequencies, while twice the imaginary part corresponds to the effective loss rates of the hybrid modes.

The nature of the coupled modes is determined by the argument of the square root. At resonance ($\omega_c=\omega_m$), the quantity under the square root becomes real only when
\begin{equation}
    4g > |\kappa_m - \kappa_c|.
    \label{Eq:SCC}
\end{equation}
In this regime, the real parts of $\omega_\pm$ split, giving rise to an avoided crossing in the spectrum. When this condition is not satisfied, the square root is purely imaginary and the coupling results only in modified linewidths without a resolvable mode splitting.

While Equation~\ref{Eq:SCC} defines the condition for observing a spectral splitting, it does not guarantee that the system operates in the strong coupling regime. A more stringent and commonly used criterion requires the coupling strength to exceed the dissipation rates of both subsystems,
\begin{equation}
    g \gtrsim \kappa_m, \kappa_c,
\end{equation}
which ensures coherent exchange of excitations between magnons and photons.\cite{cavity_optomechanics2014} In this regime, the hybrid modes are well resolved and exhibit coherent dynamics. In contrast, systems that satisfy Equation~\ref{Eq:SCC} but not this condition may display an avoided crossing without reaching the fully coherent strong coupling regime.

\section{Device A-F data and fits}

\subsection{Raw spectra and background subtraction}
\begin{figure}
    \centering
    \includegraphics[width=0.8\linewidth]{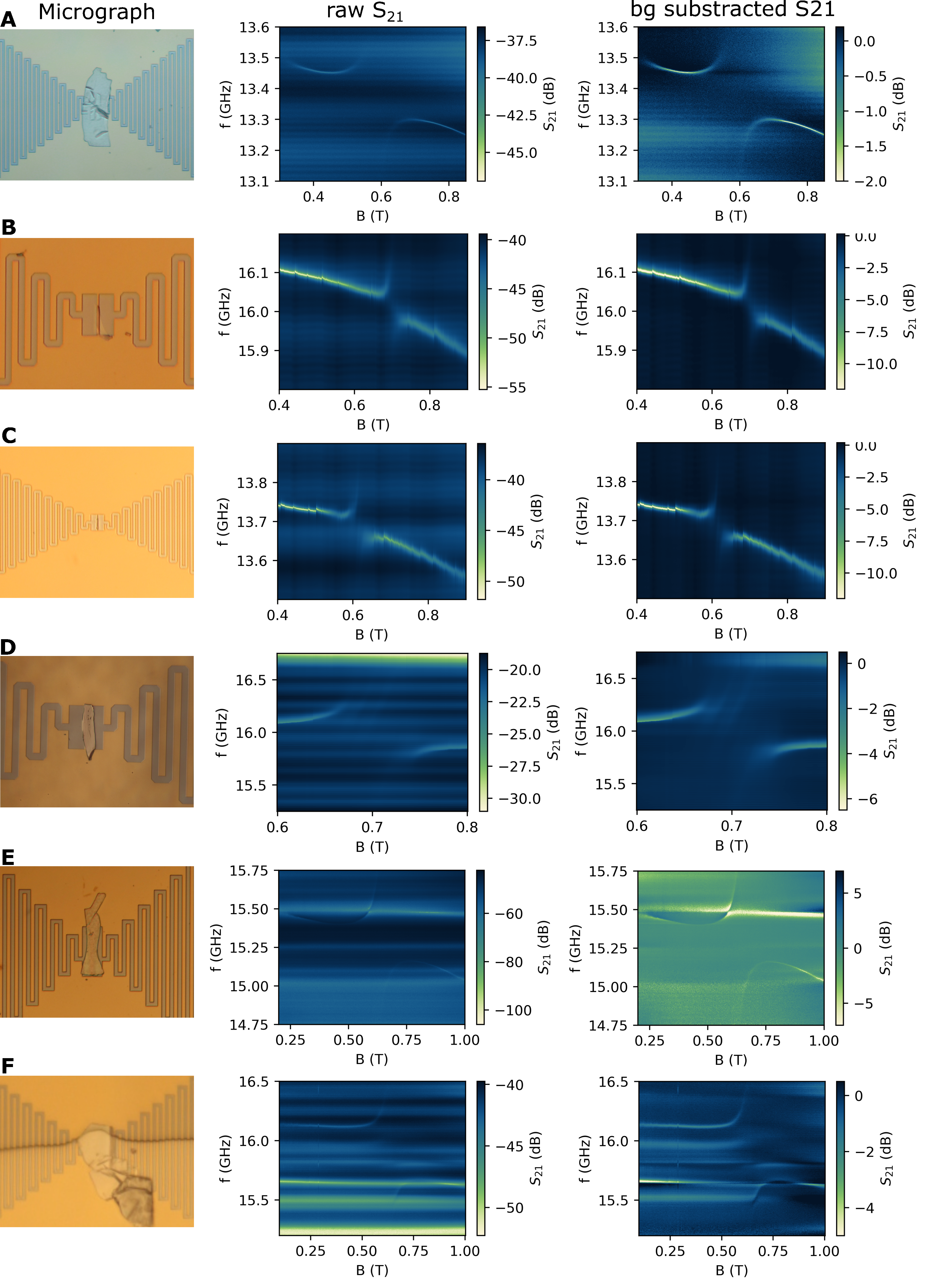}
    \caption{Each row shows, for a given device listed in Table~1 of the main text (A-F), an optical micrograph (left), the raw transmission spectrum (center), and the background-subtracted transmission spectrum (right).}
    \label{fig:data}
\end{figure}

The measured transmission spectra were first corrected to isolate the resonant features. For each magnetic field, a reference trace was selected from a field range without resonances in the frequency window of interest. Typically, traces from the opposite branch of the spectrum were used (e.g., the upper branch is corrected using lower-field traces and vice versa). This removes the slowly varying background and enhances the visibility of the resonant dips. All raw and background-subtracted spectra are shown in \textbf{Figure~\ref{fig:data}}.

\subsection{Fitting of resonance frequencies and linewidths}
\begin{figure}
    \centering
    \includegraphics[width=\linewidth]{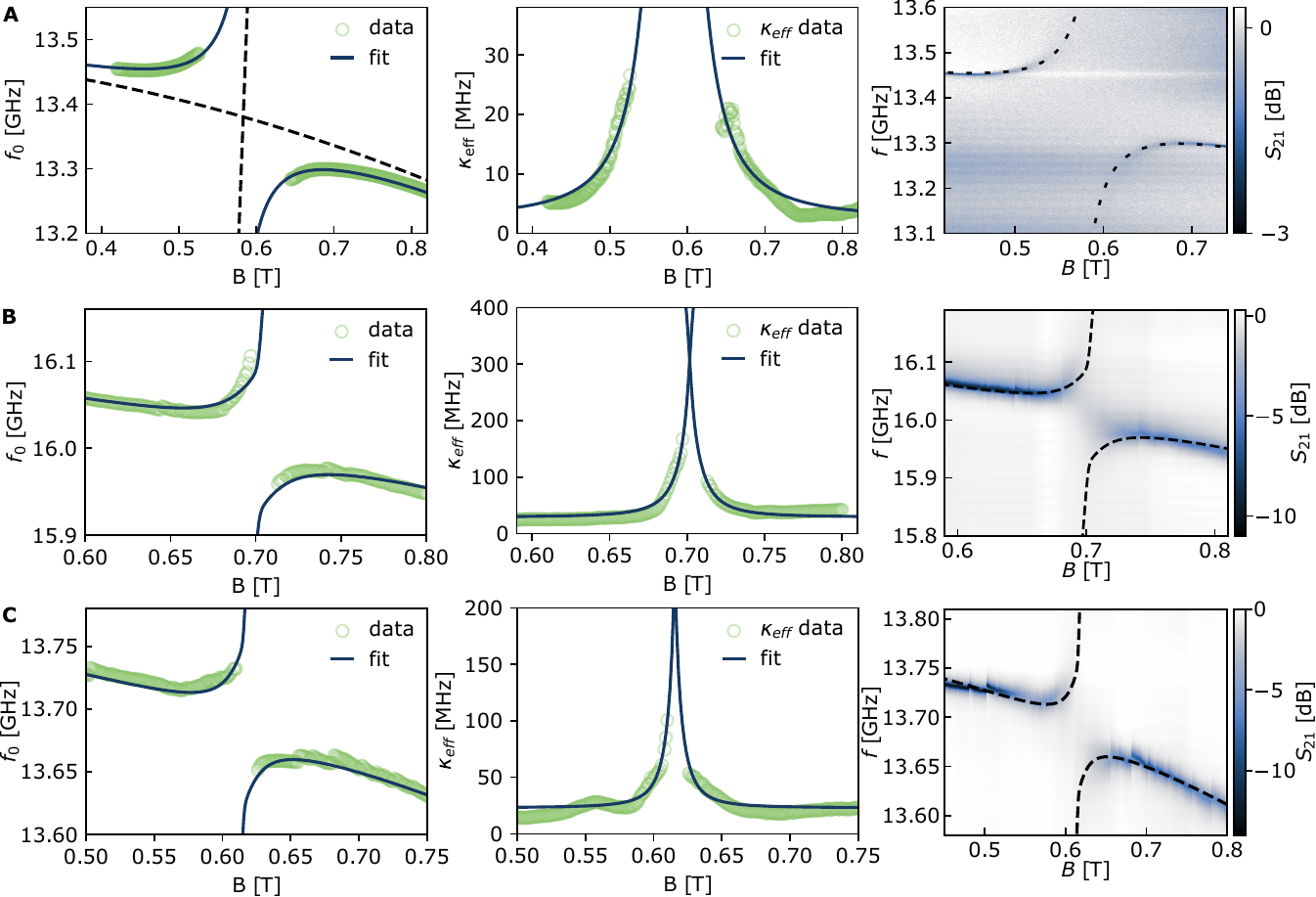}
    \caption{Fits of samples A–C. Each row corresponds to one sample. The left column shows the extracted resonance frequencies with the corresponding fit. The middle column shows the extracted linewidths with the fit, and the right column shows the frequency fit overlaid on the background-subtracted transmission map.}
    \label{fig:ABC}
\end{figure}

\begin{figure}
    \centering
    \includegraphics[width=\linewidth]{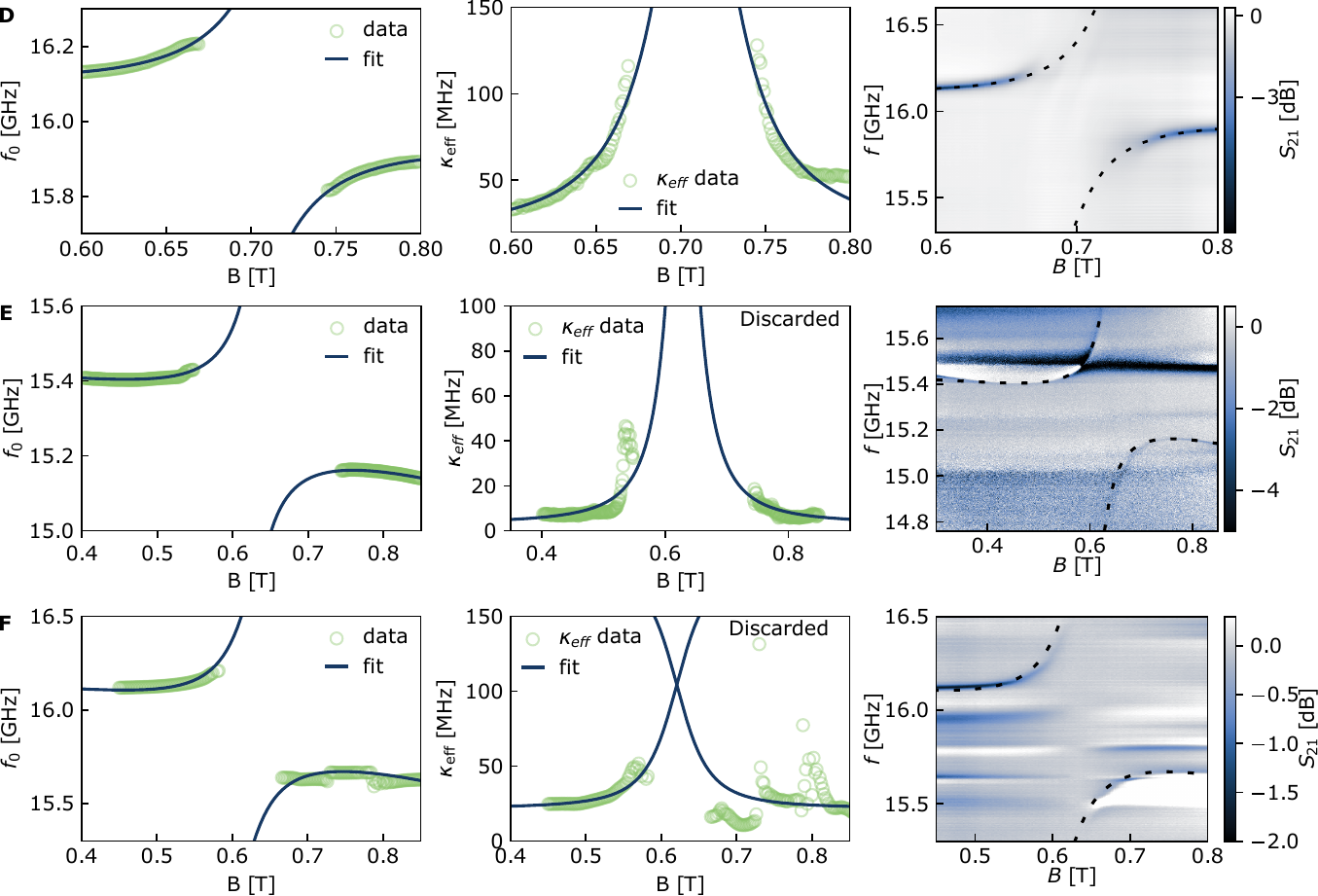}
    \caption{Fits of samples D-F. Each row corresponds to one sample. The left column shows the extracted resonance frequencies with the corresponding fit. The middle column shows the extracted linewidths with the fit, and the right column shows the frequency fit overlaid on the background-subtracted transmission map. The extracted magnon linewidths for devices E and F are not reported, as spurious RF modes in the vicinity of the resonator prevent reliable linewidth extraction.}
    \label{fig:DEF}
\end{figure}

Each background-corrected trace was fitted independently using Equation~\ref{S21} to extract the resonance frequency and total linewidth at each magnetic field. The extracted values were then simultaneously fitted to the coupled magnon–photon model (Equation~\ref{eq:omegapm}), with the magnon dispersion $\omega_m(B)$ described by Equation~\ref{eq:omegafmr} and the cavity dispersion $\omega_c(B)$ following the quadratic dependence of Equation~\ref{eq:quadratic}. Both the magnon and cavity linewidths were assumed constant across the avoided crossing, which is a good approximation for the magnons; the cavity linewidth is known to vary slightly with field. 
The fitted coupling strengths $g$ are robust and can be directly verified against the measured mode splitting, while the linewidths are more approximate due to reduced visibility near the avoided crossing. For devices E and F, spurious RF modes prevented reliable extraction of linewidths; the coupling could still be extracted. The uncertainty in the extracted parameters is estimated from the covariance matrix of the fit and is lower than 2\% for $g$ and lower than 6\% for the linewidths in all devices for which the parameters are reported. Fits and corresponding data are shown in \textbf{Figures~\ref{fig:ABC}} and \textbf{\ref{fig:DEF}}, and the data and fitting code are available in \cite{zenodo}.

\subsection{Atomic force microscopy}
\begin{figure}
    \centering
    \includegraphics[width=0.6\linewidth]{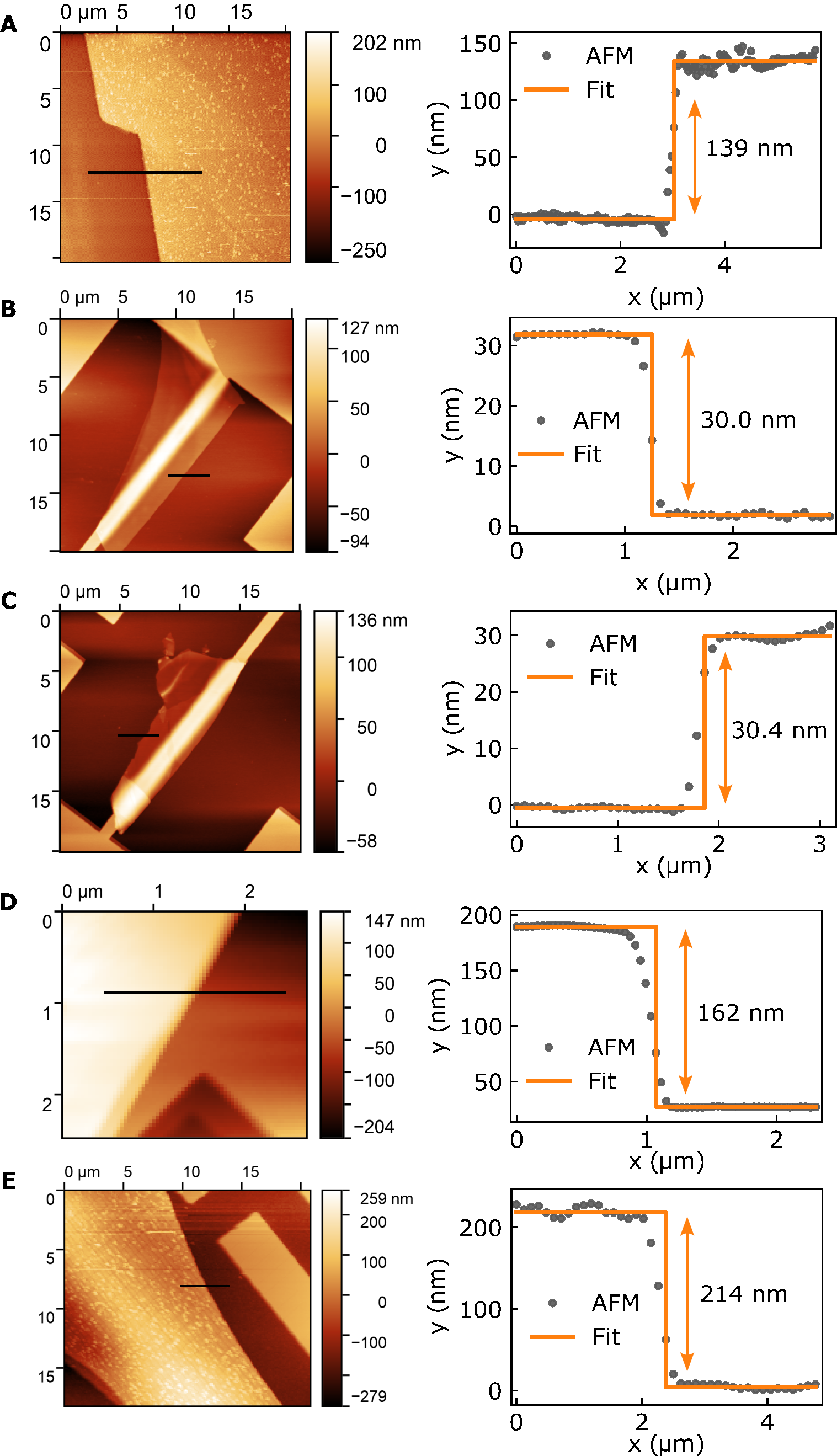}
    \caption{AFM characterization of the devices. For each device, the left panel shows the AFM image with the linecut indicated by a black line, and the right panel shows the corresponding height profile. The flake thickness is extracted by fitting to a step function.}
    \label{fig:afm}
\end{figure}

The thickness of the CGT flakes was measured using atomic force microscopy (AFM). Linecuts across the flakes were fitted to step functions to determine the flake thickness. The extracted thicknesses and linecuts are presented in \textbf{Figure~\ref{fig:afm}}.  Devices B and C exhibit stepped regions; in these cases, the thickness is taken from the plateau with the largest area. The error in the thickness determination is approximately 1 nm.

\section{Additional magnon modes in Cr$_2$Ge$_2$Te$_6$}

\begin{figure}[h]
    \centering
    \includegraphics[width=0.5\linewidth]{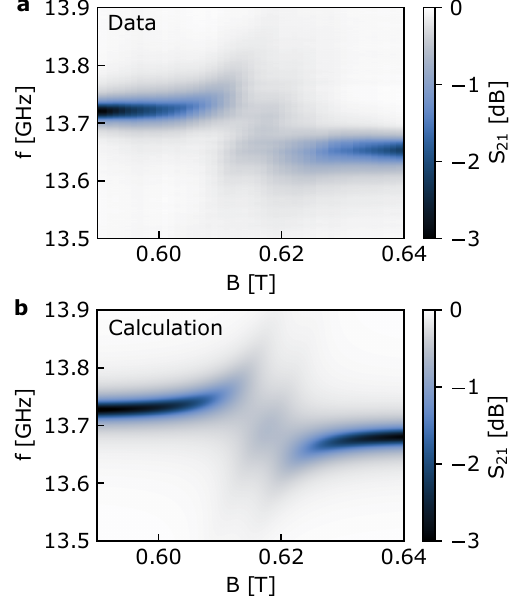}
    \caption{\textbf{a} Background-subtracted absorption spectrum of a 30~nm flake (device C) measured at 2~K as a function of in-plane magnetic field, showing an avoided crossing with an additional central feature.  
    \textbf{b} Calculation of $S_{21}$ for two magnon modes coupled to a cavity according to Equation~\ref{eq:SImodes}. Parameters: magnon frequency spacing 200~MHz, couplings 85 and 70~MHz, magnon damping 210~MHz; resonator external and internal linewidths 25 and 5~MHz.}
    \label{fig:multimode}
\end{figure}

For device C, an additional diagonal feature is observed in the center of the avoided crossing when we zoom in and enhance the colormap contrast, Figure~\ref{fig:multimode}a. To investigate whether hybridization with multiple spin-wave modes could produce such a feature, we model the system using a resonator coupled to several magnon modes \cite{mandal_coplanar_2020}:  

\begin{align}\label{eq:SImodes}
S_{21}(\omega)
= 1 - 
\frac{\kappa_e}{
\, i(\omega - \omega_c)
+ \dfrac{\kappa_e + \kappa_i}{2}
+ \displaystyle\sum_{n=0}^{N}
    \frac{g_n^{2}}{
        i(\omega - \omega_{\mathrm{fmr},n})
        + \dfrac{\kappa_{m,n}}{2}
    }
}.
\end{align}

Here, the FMR mode frequencies are assumed to have the same magnetic-field dependence as the Kittel mode, with a fixed frequency offset. The calculated spectrum for two magnon modes is shown in Figure~\ref{fig:multimode}b, capturing the main features of the data.  

 The thickness sets the wavevector quantization, while the relatively weak interlayer exchange stiffness in van der Waals magnets compared to systems such as YIG \cite{Vervelaki2024, Zhu2021, Klingler_2015} leads to small mode spacings even in thin layers.

The additional diagonal feature indicates the presence of multiple magnon modes, consistent with prior reports in CGT \cite{Zollitsch2023} and other van der Waals magnets \cite{kapoor2021}. These modes may correspond to standing spin-wave modes along the flake thickness. In thin magnetic films, boundary conditions along the thickness lead to quantized exchange modes. The spacing between these modes depends on the flake thickness and the interlayer exchange stiffness. The thickness sets the wavevector quantization while the relatively weak interlayer exchange stiffness in van der Waals magnets compared to systems such as YIG \cite{Vervelaki2024, Zhu2021, Klingler_2015} leads to small mode spacings even in thin layers. As a result, flakes with thicknesses of only tens of nanometers can have mode separations of tens to hundreds of MHz. 

The localized magnetic field generated by the resonator inductor relaxes the usual selection rules for uniform excitation, enabling coupling to higher-order spin-wave modes with finite wavevector. When the frequency separation is small compared to the intrinsic linewidths, the spectral contributions overlap, producing an apparent linewidth that is larger than that of a single mode. Interface properties, such as surface oxidation, may further affect the effective mode spacing \cite{Portis1963}. Understanding and controlling these modes remains an open challenge and could provide a route to probe spin-wave dynamics in van der Waals magnets approaching the two-dimensional limit.

\subsection{Estimation of standing spin-wave mode spacing}

With Equation~\ref{eq:omegafmr}, we considered only the uniform magnetisation mode, neglecting higher-order modes. For non-uniform modes, both magnetostatic and exchange energies modify the spectrum. In general, all contributions should be included; however, here we show that, even neglecting the magnetostatic energy, the exchange interaction alone yields mode spacings of the same order as those hinted at experimentally.

The model of Equation~\ref{eq:omegafmr} can be extended by including an exchange contribution~\cite{kalinikos_theory_1986}
\begin{equation}\label{eq:omegaexchange}
  \omega_{\text{m}, n} = \mu_0\gamma\sqrt{(H_\text{dc}+M_s(\lambda_{\text{ex}}k_n)^2)(H_\text{dc}+M_s(\lambda_{\text{ex}}k_n)^2+M_s - H_\text{k})}.
\end{equation}
Here, $\lambda_\text{ex}$ is the exchange length and $k_n$ is the wavenumber of the mode. Approximating the magnetic flake as a rectangular three-dimensional box, the allowed wavevectors are given by
\begin{equation}\label{eq:kn}
    k_n^2 = k_x^2 + k_y^2 + k_z^2,
\end{equation}
with
\begin{equation}\label{eq:kx}
    k_i = \frac{\pi n_i}{L_i},
\end{equation}
where $n_i = 0,1,2,\dots$ and $L_i$ is the flake dimension along direction $i$. This expression assumes pinned boundary conditions at the sample surfaces, which provide an upper bound for the mode spacing. The $k_n^{-2}$ dependence implies that reducing the lateral dimensions increases the mode spacing. We consider two possibilities. First, out-of-plane standing spin-wave modes, in which the spin oscillation amplitude varies sinusoidally along the thickness of the flake (with wavevector $k_n$) while remaining uniform in the lateral directions. Second, in-plane modes, for which the oscillation amplitude is uniform across the thickness but varies along one of the lateral directions of the flake.

For CGT, the intralayer exchange stiffness is $A_\text{ex} = 1.58$~pJ\,m$^{-1}$, corresponding to an exchange length $\lambda_\text{ex} =\sqrt{\frac{2A_\text{ex}}{\mu_0 M_s^2} } = 9.42$ nm. Along the out-of-plane direction, the exchange interaction is significantly weaker, resulting in an exchange stiffness of approximately $3\%$ of the intralayer value and a correspondingly reduced exchange length. \cite{Vervelaki2024}

\begin{figure}[h]
    \centering
    \includegraphics[width=0.8\linewidth]{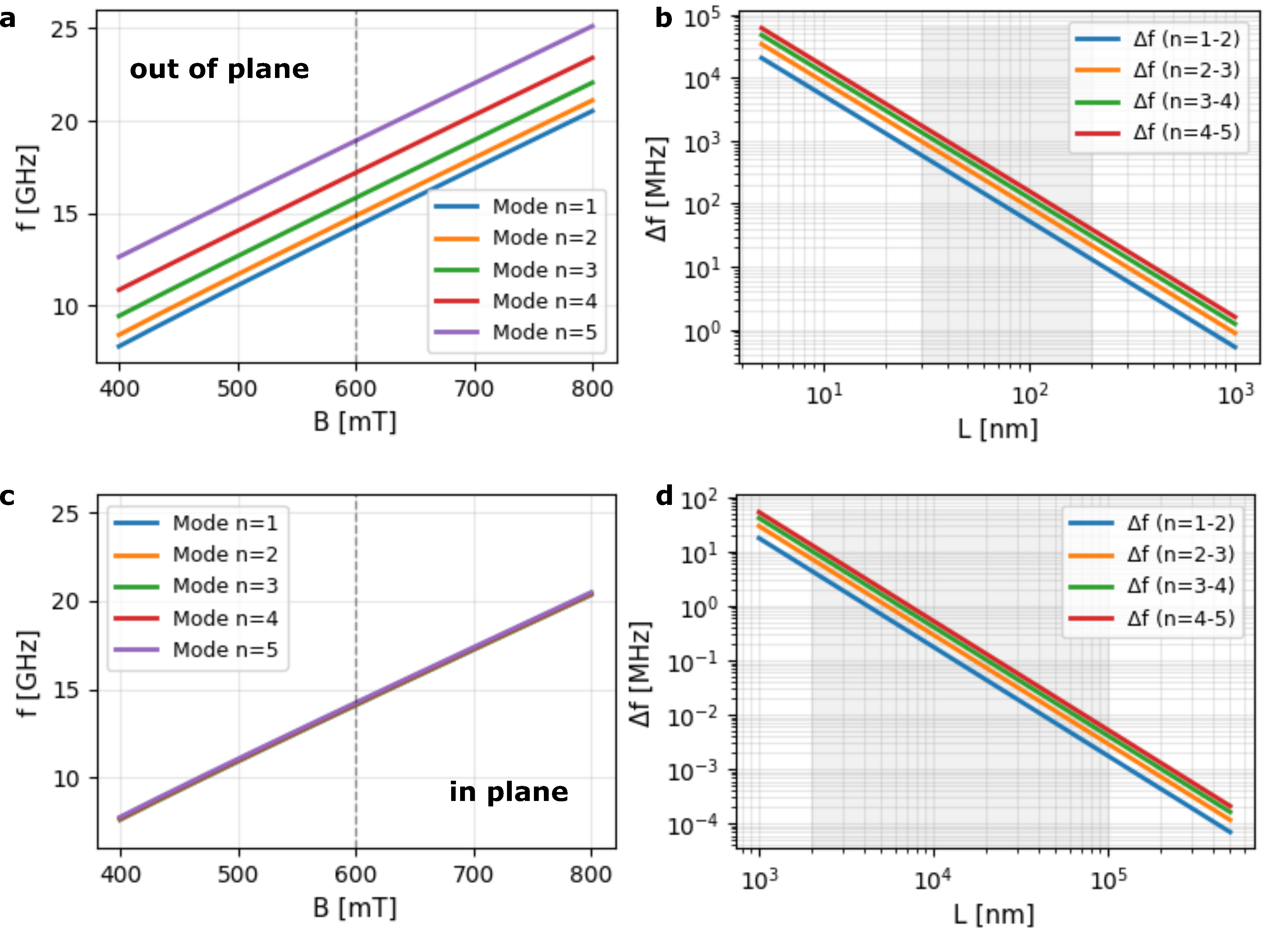}
    \caption{
    Exchange-dominated standing spin-wave modes in a confined CGT flake according to Equation \ref{eq:omegaexchange}. 
    (a) Calculated resonance frequencies for out-of-plane modes as a function of magnetic field. 
    (b) Corresponding mode spacing between consecutive modes for the out-of-plane case up to the 5th mode. 
    (c) Calculated resonance frequencies for in-plane modes. 
    (d) Mode spacing for the in-plane case up to the 5th mode. 
    }
    \label{fig:exchange}
\end{figure}

Using these parameters, we compute the mode spectrum for both out-of-plane and in-plane standing spin-wave modes. The out-of-plane spectrum is shown in Figure~\ref{fig:exchange}a, with the corresponding mode separation as a function of thickness in panel~b. Panels~c and~d show the equivalent results for in-plane modes, where the distance is now the lateral size. For a 30~nm thick flake, the mode separation is on the order of hundreds of MHz, comparable to the magnon linewidth. In contrast, for a $1~\mu$m-wide flake, the mode spacing falls below 1~MHz, well below the linewidth. These calculations point to thickness-confined modes as the origin of the observed broad linewidth and closely spaced spectral features, while ruling out any contribution of exchange for in-plane modes. If this is the case, then going to thinner samples should lead to well-separated modes with the intrinsic linewidth of CGT. We note that the estimated mode spacing depends on the assumed exchange stiffness and boundary conditions; however, variations within realistic parameter ranges do not change the order of magnitude of the results. Additionally, surface effects, such as oxidation, can further reduce the mode spacing~\cite{Portis1963}, an effect that has been reported in standing spin-wave modes in van der Waals magnets~\cite{kapoor2021}.

\FloatBarrier
\bibliographystyle{MSP}
\bibliography{SI}